\documentclass[12pt,a4paper]{article}
\usepackage{graphicx}
\usepackage{epsfig}
\usepackage{subfig}
\usepackage{amssymb}
\begin{document}

\title {New measure of multifractality and its application in finances}

\author {Dariusz Grech\footnote{dgrech@ift.uni.wroc.pl} and Grzegorz Pamu{\l}a\footnote{gpamula@ift.uni.wroc.pl} \\
Institute of Theoretical Physics\\
Pl.
M. Borna 9, University of Wroc{\l}aw, PL-50-204 Wroc{\l}aw, Poland}
\date{}

\maketitle
\begin{abstract}
We provide an alternative method for analysis of multifractal properties of time series.
The new approach takes into account the behavior of the whole multifractal profile of the generalized Hurst exponent $h(q)$ for all moment orders $q$, not limited only to the edge values of $h(q)$ describing in MFDFA scaling properties of smallest and largest fluctuations in signal.
The meaning of this new measure is clarified and its properties are investigated for synthetic multifractal data as well as for real signals from stock markets.
We show that the proposed new measure is free of problems one can meet in real nonstationary signals, while searching their multifractal signatures.
\end{abstract}

$$
$$
\\
\section{Introduction}

The multifractal detrended fluctuation analysis (MFDFA)  \cite{MF-DFA} became in last ten years the major technique used for studying the multifractal properties in complex systems and in time series. More then 500 papers discusses MFDFA issues in variety of problems related with complexity (see, e.g.,
\cite{MF-seismology_1}--\cite{czarnecki}).
The central role in MFDFA is played by the $q$-th moment of fluctuation function ($q\in \textsc{R}$) defined as \cite{MF-DFA}

\begin{equation}
F(q,\tau)=\left\{\frac{1}{2N}\sum^{2N}_{k=1} [\hat{F}^2(\tau,k)]^{q/2}\right\}^{1/q}
\end{equation}

for $q\neq 0$, and
\begin{equation}
F_0(\tau)=\exp\left\{\frac{1}{4N}\sum^{2N}_{k=1} \ln[\hat{F}^2(\tau,k)]\right\}
\end{equation}
for $q=0$, where
\begin{equation}
\hat{F}^2(\tau,k)=\frac{1}{\tau}\sum^{\tau}_{j=1}\left\{x_{(k-1)\tau+j}-P_k(j)\right\}^2
\end{equation}
and $x_j$ ($j=1,\ldots,N\tau$) are data in series, $\tau$ is the size of window box in which detrending is performed, while $P_k(j)$ is the polynomial trend subtracted for $j$-th data in $k$-th window box ($k=1,\ldots,N$).

The observed scaling law
\begin{equation}
F(q,\tau)\sim \tau^{h(q)}
\end{equation}
is crucial within MFDFA to estimate the multifractal properties of a signal given by the profile of generalized Hurst exponents $h(q)$.
The strength of multifractality present in data is usually studied as a spread $\Delta h$

\begin{equation}
\Delta h \equiv h(-Q) - h(Q)
\end{equation}
of some edge values of generalized Hurst exponent $h^{\mp}$  calculated at the fixed negative ($-Q$) and positive ($Q$) moments respectively ($Q>0$).

In the case of stationary series, $h(q)$ is proven to be a monotonically decreasing function and therefore $\Delta h>0$ \cite{Kantelhardt-arxiv}.
However, many examples shown recently in literature for synthetic data \cite{spur-corr-MF, zunino1} as well as for realistic signals \cite{czarnecki,FENS} reveal that multifractal properties may lay far outside this simplest scenario and the spread $\Delta h $, defined as the difference of generalized Hurst exponents calculated for small negative  and large positive fluctuation moments, is not indicative in these cases.

A non-monotonic behavior of $h(q)$ profile is observed for instance in nonstationary data what leads to twisted $f(\alpha)$ multifractal spectrum \cite{czarnecki} if the H\"{o}lder description of multifractality is adopted with a help of Legendre transform \cite{legendre1,legendre2}.
In the case of nonstationary data with periodicity, white or color noise added (see, e.g., \cite{spur-corr-MF,zunino1}) one may see domains where $h(q)$ is either increasing or decreasing with $q$, local maxima in $h(q)$ are formed or even $h(q\rightarrow{-\infty})<h(q\rightarrow{+\infty})$ suggesting that big fluctuations may appear more often than small ones. It is contrary to observations in stationary data \cite{Kantelhardt-arxiv} where it should be the other way round. There are examples of realistic time series where the multifractal profile, despite small and insignificant  $\Delta h$ spread, significantly expands in the central part for moment orders $|q|<Q$. One finds such behavior for instance for nonlinearly transformed financial data containing extreme events (see, e.g., \cite{FENS}). The similar spread $\Delta h$ may correspond also to significantly different characteristics of the $h(q)$ profile in the center for $|q|<Q$.

The above results convince that a traditional measurement of multifractality within MFDFA, based entirely on the $\Delta h$ spread may be misleading.
Therefore we propose in this article an alternative measure of multifractal properties
which takes into account the behavior of the multifractal profile $h(q)$ as a whole for all $q$-th order moments, not limited only to the edge values $h^{\mp}$ found at $\mp Q$.

\section{Introducing new measure}

As pointed above, the new measure should be sensitive to behavior of the whole $h(q)$ profile in the wide range of $q$ moments ($-Q\leq q\leq Q$) and should not neglect the multiscaling properties within this range of $q$'s, contrary to the standard measure based on the spread $\Delta h$ alone which takes into account just the edge values of the generalized Hurst exponents $h(\mp Q)$.

In the simplest case of multifractal data with the abandoned effects of so called multifractal bias, generally  discussed in \cite{PhysicaA2013}, a natural extension of $\Delta h$ can be defined as a cumulated distance between the multifractal profile $h(q)$ and the main Hurst exponent value $H\equiv h(2)$ for considered series.

One may write then a new measure $\Delta h^{(2)}_h$ in the integral form
\begin{equation}
\Delta^{(2)}_h = \frac{1}{Q}\int_{-Q}^{Q}|h(q)-h(2)|\,\mathrm{d}q.
\end{equation}

It turns out that $\Delta^{(2)}_h$ has an interesting interpretation of the scaling exponent in the new power law linking the geometric mean of detrended $q$-th order fluctuations $F(q,\tau)$ of the signal for all $q$ values with the size $\tau$ of the time window.
Indeed, consider the  normalized detrended $q$-th moment of fluctuation $\tilde{F}(q,\tau)$ defined as follows
\begin{equation}
\tilde{F}(q,\tau) =\left\{
\begin{array}{c l}
\frac{F(q,\tau)}{\sigma_{\tau}} & \mathrm{for\ } q\leq 2\\
\frac{\sigma_{\tau}}{F(q,\tau)}& \mathrm{for\ } q>2
\end{array}\right.
\end{equation}
where $\sigma_{\tau}$ is the standard deviation of detrended fluctuation ($F(2,\tau)$) in time window of length $\tau$.
Using the power law from Eq.(4) we may write the relationship
\begin{equation}
|h(q)-h(2)|\ln (\tau) = \ln \tilde{F}(q,\tau) + C(q)
\end{equation}
where $C(q)$ is some function generally dependent on $q$ but independent on $\tau$. Hence, using definition from Eq.(6)

\begin{equation}
Q \Delta^{(2)}_h\ln (\tau) = \int_{-Q}^{Q}\ln \tilde{F}(q,\tau)\,\mathrm{d}q + \int_{-Q}^{Q}C(q)\,\mathrm{d}q
\end{equation}
Replacing $\mathrm{d}q$ by $2Q/n$ and $q$ by $-Q+k\frac{2Q}{n}$, where $k=0,1,...,n$, the first integral in RHS of Eq.(9) can be approximated by a discrete sum
\begin{equation}
\sum_{k=0}^n\ln \tilde{F}(-Q+k\frac{2Q}{n},\tau) \frac{2Q}{n} = Q \ln \left [\prod_{k=0}^n \tilde{F^2}(-Q+k\frac{2Q}{n},\tau)\right]^{\frac{1}{n}}
\end{equation}
Comparing Eqs(9) and (10), one gets  in the continuous limit $n\rightarrow\infty$
\begin{equation}
Q \Delta^{(2)}_h \ln \tau = \lim_{n\rightarrow\infty}Q\ln\left[\prod_{k=0}^n \tilde F^2(-Q+k\frac{2Q}{n}, \tau)\right]^{\frac{1}{n}} + \int_{-Q}^{Q}C(q)\,\mathrm{d}q
\end{equation}
The latter relation  builds after simple rearrangement the postulated power law
\begin{equation}
\lim_{n\rightarrow\infty}\left[\prod_{k=0}^n \tilde F^2(-Q+k\frac{2Q}{n}, \tau)\right]^{\frac{1}{n}} \sim \tau^{\Delta^{(2)}_h}
\end{equation}
Note that the standard spread of generalized Hurst exponents values used in MFDFA to determine the multifractal content can also be interpreted in the above context as the scaling exponent in the power law
\begin{equation}
\frac{F(-Q,\tau)}{F(Q,\tau)} \sim \tau^{\Delta h}
\end{equation}
which links together $\tau$ with the ratio of the smallest ($F(-Q,\tau)$) and the biggest ($F(Q,\tau)$) detrended $q$ deformed fluctuations. Hence, the scaling in Eq.(12) related with geometric mean of all detrended fluctuations $F^2(q,\tau)$, can be regarded as the modification of MFDFA method we call GMFDFA (geometric mean MFDFA).

\begin{figure}[p]
\centering
\includegraphics[width=11truecm]{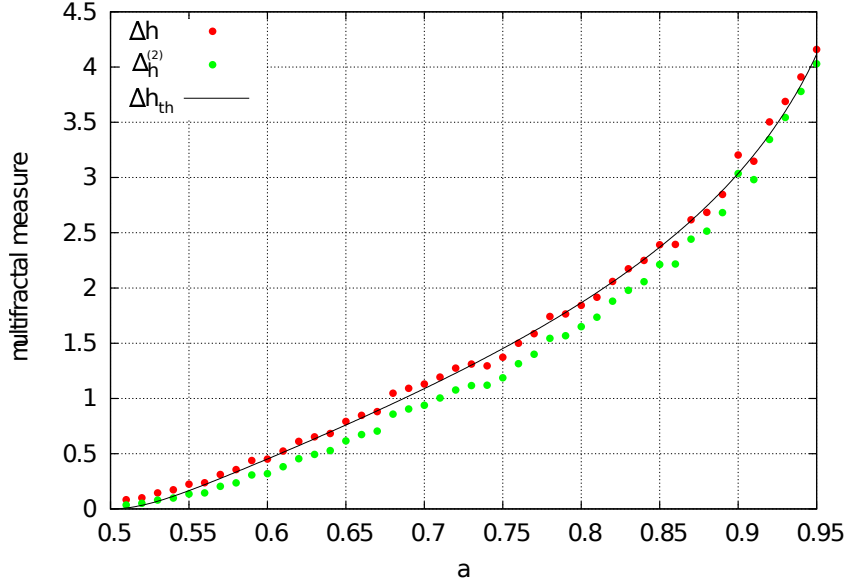}
\caption{Comparison of standard multifractal measures $\Delta h$ (simulated), $\Delta h_{th}$ (theoretically estimated from Eq. (14)) and the newly proposed $\Delta^{(2)}_h$ shown for synthetic series of length $L=2^{16}$ generated from random binomial cascade algorithm for different values of parameter $a$}
\label{Delta_vs_a}
\end{figure}

We will test the behavior of the new measure analyzing first the synthetic multifractal signals with apriori well known multiscaling properties. They were generated from the binomial random cascade algorithm \cite{cascades} for the almost complete range of $a$ parameter describing the richness of multifractal content in synthetic signal ($0.5<a\leq0.95$) with a step $\Delta a=0.01$. Fig.~\ref{Delta_vs_a} shows the plot of analytical expectation $\Delta h_{th}(q)=h(-q)-h(q)$ in this model  obtained from the relation \cite{cascades2}
\begin{equation}
h(q) = \left\{\begin{array}{l l}
\frac{1}{q}\left(1-\log_2(a^q+(1-a)^q)\right) & \mathrm{for\ }q\neq0\\
-\frac{1}{2}(\log_2a+\log_2(1-a)) & \mathrm{for\ }q=0
\end{array}\right.
\end{equation}
and compared with two numerical findings, respectively for the standard $\Delta h$ and "new" $\Delta_h^{(2)}$ measure, calculated for $Q=15$ from numerical simulation of $h(q)$ profile.
These plots prove that in a case of synthetic data new measure of multifractal content is qualitatively and quantitatively indistinguishable from the standard measure based on the spread  $\Delta h$ of the edges $h(\pm Q)$.

The simple recipe for the new measure will have to be modified if so called multifractal bias  effects are to be included in analysis \cite{FENS,PhysicaA2013,arxiv}. Let us recall here main issues connected with such multifractal bias.
It is defined as the continuous set of profiles $h(q)$ calculated for a given system or signal, which do not correspond to multifractal properties of this system at some confidence level, despite $h(q)\neq const$.
Such bias can be seen particularly in short signals where accidental fluctuations, not related with multifractal properties, contribute to fluctuation function and dominate  over fluctuations having their origin in multifractal properties of the signal due to small statistics of short data. It raises the $h(q<0)$ edge of the multifractal spread. On the other hand, the influence of large fluctuations in short data set can be suppressed since MFDFA algorithm may treat such fluctuations as longer trends and eliminate them. In both cases the observed $\Delta h$ spread is increased comparing with the corresponding value obtained for much longer data and grows with persistency in signal \cite{PhysicaA2013}.
Another source of multifractal bias are nonlinear transformations performed on data which introduce always additional registered multifractal effect. 

The level of multifractal bias in series can be calculated or simulated with a number of semi-analytic formulas provided in \cite{PhysicaA2013}. Note that deliberately we do not mix this bias effect here with so called spurious multifractality \cite{spur-corr-MF} caused by various external contaminations in signal or nonstationarities coming, e.g., from the influence of noise present in data.

\begin{figure}[p]
\centering
\includegraphics[width=11truecm]{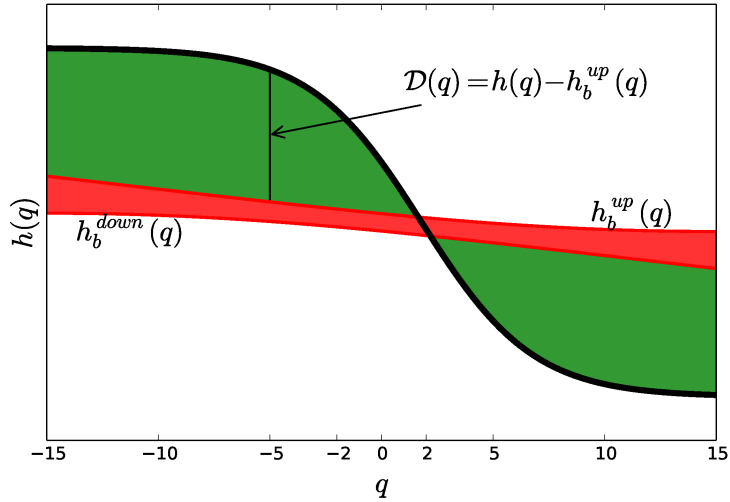}
\caption{Illustrative example of generalized Hurst exponent profile $h(q)$ with marked multifractal bias area (red) and its thresholds $h_b^{up/down}(q)$. The green area represents the cumulated distance between $h(q)$ and the multifractal bias (see Eq. (16))}
\label{artistic}
\end{figure}

When taking multifractal bias into account, one has to generalize the formula in Eq.(6), replacing $H=h(2)$ in there by the corresponding multifractal bias thresholds $h_b^{up/down}$ at the assumed confidence level ($95\%$ throughout this article). Here $h_b^{up/down}(q)$ are respectively the upper ($h_b^{up}$) and the lower  ($h_b^{down}$) edge of the  bias area (ribbon) explained in Fig.~\ref{artistic} (and shown also in Figs. \ref{cascades_withB},\ref{SP500}--\ref{NASDAQ}). Simultaneously, $h(q)$ profile in Eq. (6) has to be replaced by the observed (registered) values of the multifractal profile $h_{obs}(q)$ coming from direct application of Eq.(4) in MFDFA.

Let us define a distance function $\mathcal{D}(q)$ as

\begin{equation}
\mathcal{D}(q)=\left\{
\begin{array}{c l}
h_{obs}(q)-h_b^{up}(q) & \mathrm{for\ } h_{obs}(q)>h_b^{up}(q)\\
h_b^{down}(q)-h_{obs}(q) & \mathrm{for\ } h_{obs}(q)<h_b^{down}(q)\\
0 & \mathrm{for\ } h_b^{down}(q)\leq h_{obs}(q)\leq h_b^{up}(q)
\end{array}\right.
\end{equation}
The modified measure $\Delta_h$ in the case of multifractal bias present in signal can be taken as the average distance  $\mathcal{D}(q)$
\begin{equation}
\Delta_h = \frac{1}{Q}\int_{-Q}^{Q}\mathcal{D}(q)\,\mathrm{d}q
\end{equation}
and is presented as a green area in Fig.~\ref{artistic}.

A new measure of multifractality $\Delta_h$ generalizes the multifractal spread $\Delta h$ in such a way, that it can be used for particular profiles $h_{obs}(q)$ even if the standard $\Delta h$ measure could not be applied (e.g., when $h(-Q)<h(+Q)$ or  $\Delta h=0$). One always gets $\Delta_h\geq0$ describing multiscaling properties in a system for the whole range of fluctuation moments. The result  $\Delta_h=0$ means that the observed multifractal profile entirely lies in a biased  ribbon domain for all moment orders $q$ and hence there are no expected real multifractal effects even if $\Delta h_{obs}>0$.

Note, however, that the maximum value of $Q$ at which the new measure is calculated can not be too high, since then the contribution to the profile $h(q)$ with high volatility may be dominated by domains where $h(q)\approx\mathit{const}$.

\section{Application to synthetic and real time series}

\begin{figure}[p]
\centering
$$\quad\quad a=0.65\quad\quad\quad\quad\quad\quad\quad\quad\quad\quad a=0.75$$
\includegraphics[width=11truecm]{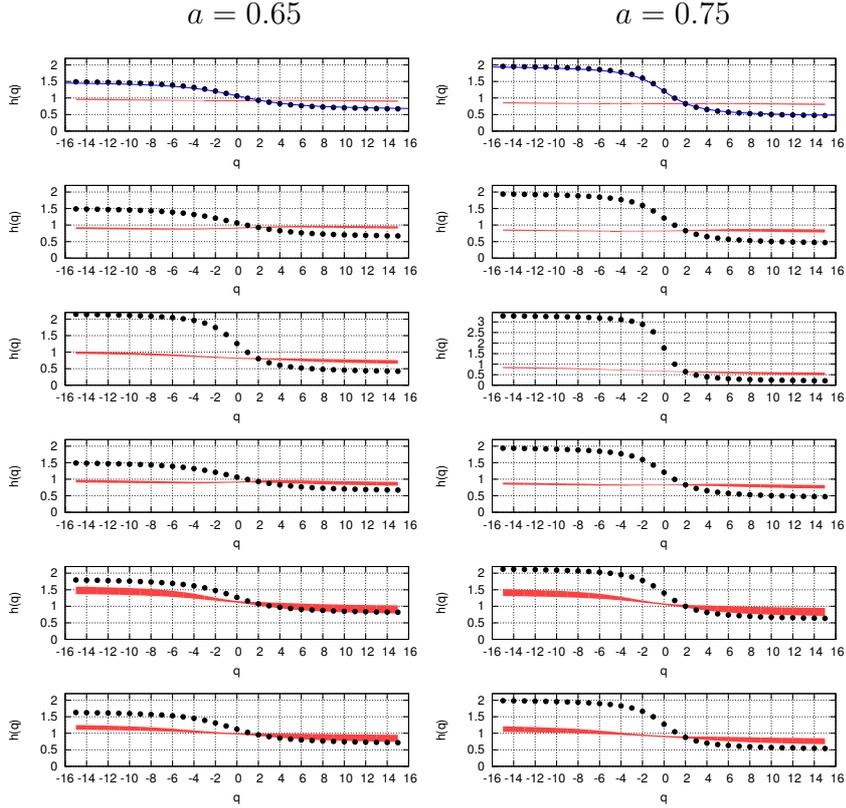}
\caption{Multifractal profiles $h(q)$ (black dots) and the corresponding multifractal bias (red area) calculated for synthetic series generated from random binomial cascade algorithm for $a=0.65$ and $a=0.75$
First row shows result for primary data in series $(\Delta x_i)$, while consecutive rows present the results of following nonlinear transformations: $|\Delta x_i|$, $(\Delta x_i)^2$, $|r_i|$, $\mu_i$, $v_i$.}
\label{cascades_withB}
\end{figure}

\begin{figure}[p]
\centering
\includegraphics[width=6truecm]{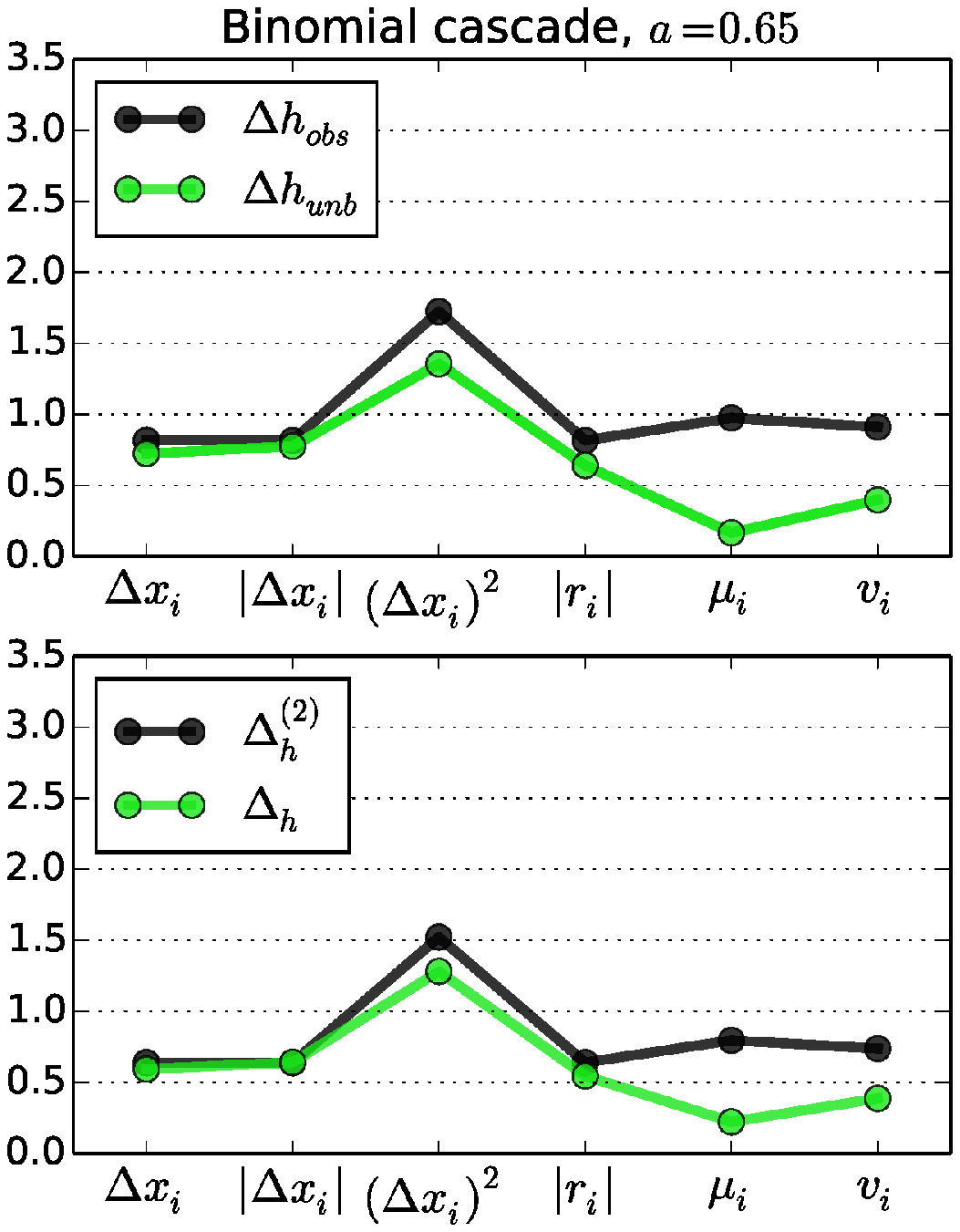}
\includegraphics[width=6truecm]{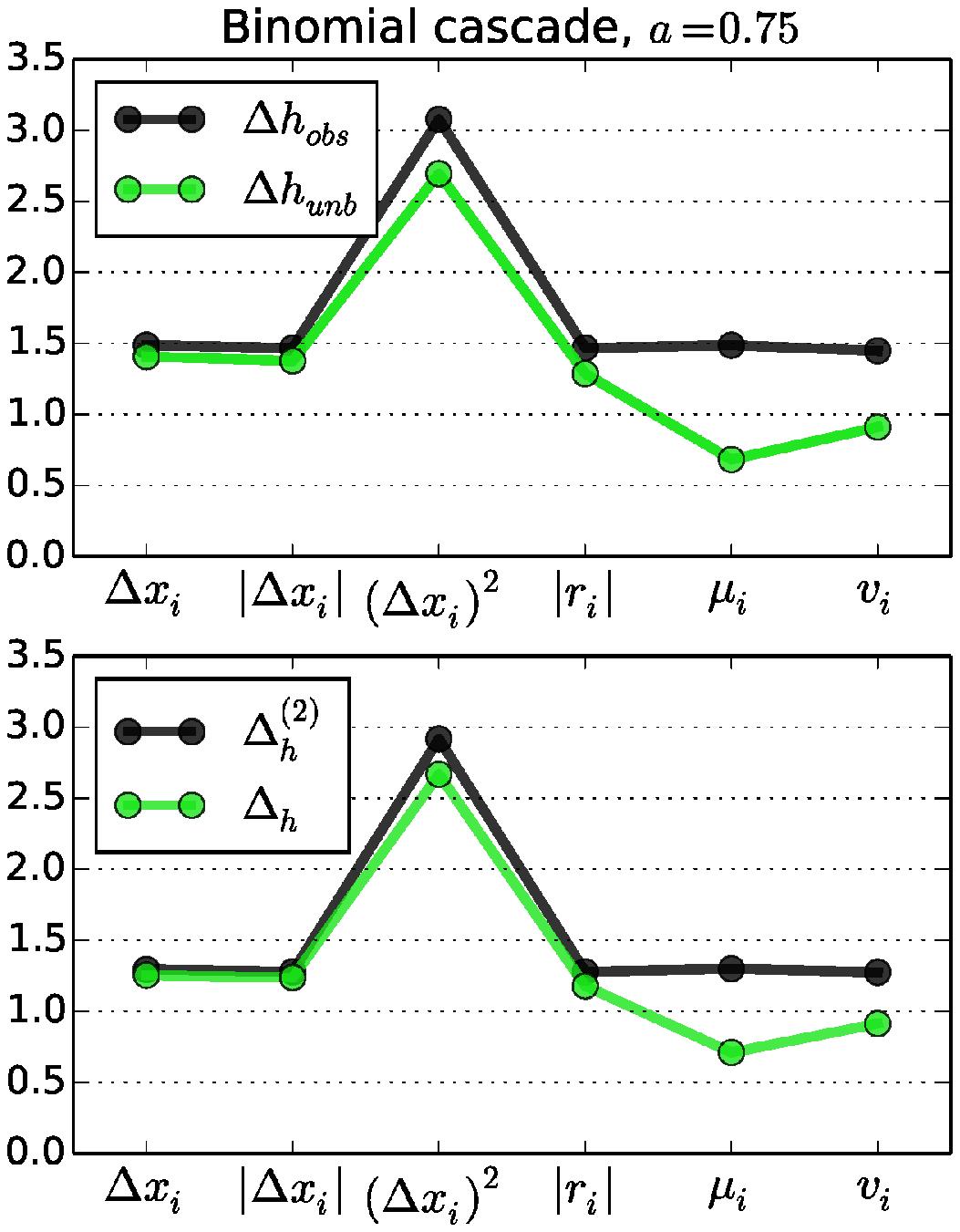}
\caption{Comparative study of multifractal measures for synthetic data generated from random binomial cascade algorithm.
Different nonlinear transformations are indicated.
The observed (biased) results are marked with black color.
Results after subtraction of multifractal bias (unbiased measure) are marked in green.
Two bottom panels show results of description with the newly proposed measure $\Delta_h$ and its biased partner $\Delta^{(2)}_h$.
}
\label{cascades-barplots}
\end{figure}

\begin{figure}[p]
\centering
\includegraphics[width=13truecm]{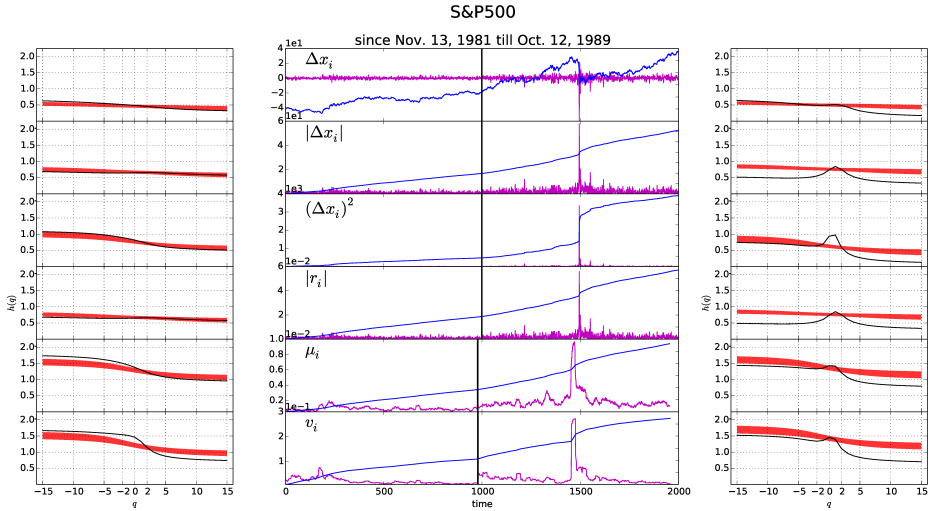}
\caption{Multifractal analysis of S\&P500 signal in adjacent time windows in "quiet" period (Nov. 13, 1981--Oct. 29, 1985) and dynamical one (Oct. 30, 1985--Oct. 12, 1989).
The $h(q)$ profiles are placed correspondingly in left and right panels
for primary and transformed data.
Each profile is supplemented by related multifractal bias (red).
Values of series increments ale marked in magenta color.}
\label{SP500}
\end{figure}

\begin{figure}[p]
\centering
\includegraphics[width=13truecm]{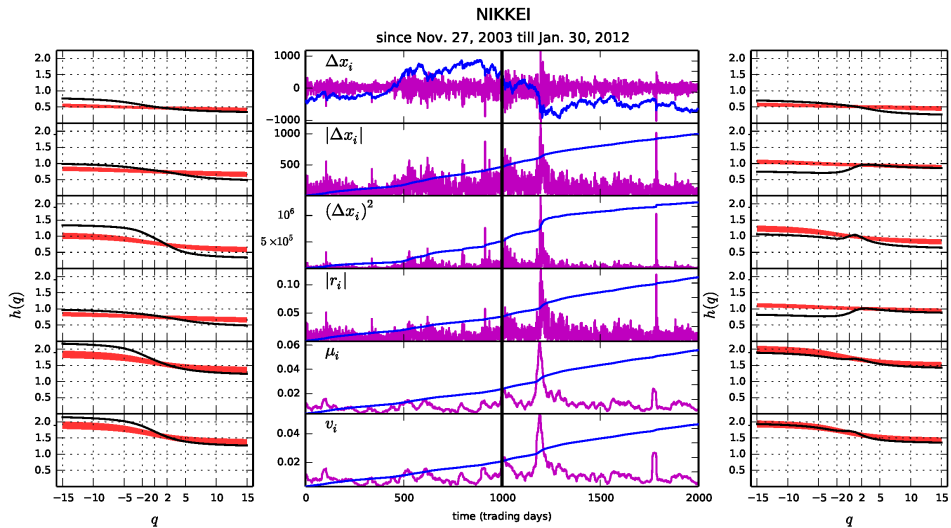}
\caption{As in Fig.~\ref{SP500}, but for Nikkei225 index with "quiet" period Nov. 27, 2003--Dec. 19, 2007 (left) and dynamical period Dec. 20, 2007--Jan. 30, 2012 (right).
}
\label{Nikkei225}
\end{figure}

\begin{figure}[p]
\centering
\includegraphics[width=13truecm]{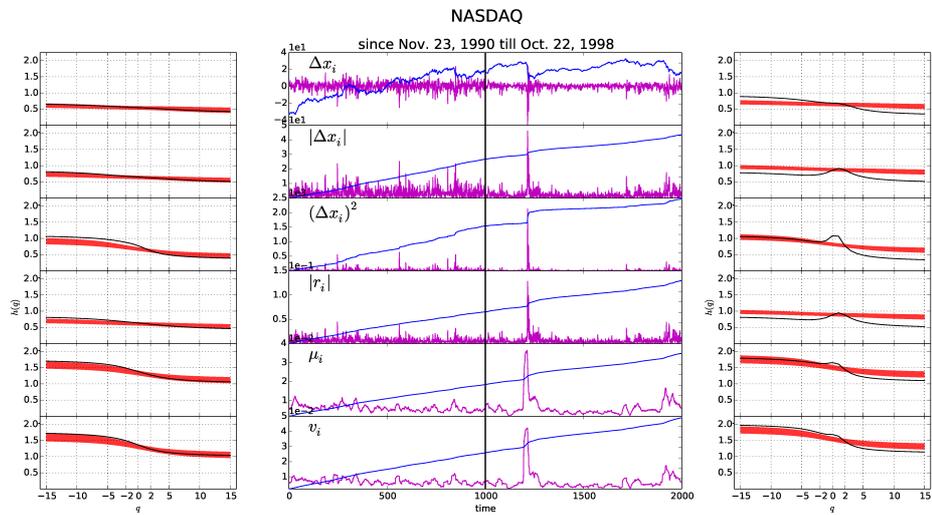}
\caption{As in Fig.~\ref{SP500}, but for NASDAQ index with "quiet" period Nov. 23, 1990--Nov. 7, 1994 (left) and dynamical period Nov. 8, 1994--Oct. 22, 1998 (right).
}
\label{NASDAQ}
\end{figure}

In order to present the qualities of proposed measure $\Delta_h$ let us start with analysis of synthetic signals obtained from binomial cascade model, additionally, taking into account the multifractal bias.
The calculated multifractal profiles $h(q)$, together with ribbon-like areas representing the multifractal bias are gathered in Fig.~\ref{cascades_withB} for two values of cascade parameter $a=0.65$ and $a=0.75$ ($L=2^{16}$).
The values obtained for three measures applied to examples from Fig.~\ref{cascades_withB}: observed (biased) spread $\Delta h_{obs}$, unbiased spread $\Delta h_{unb}$ and proposed measure $\Delta_h$ are presented in Fig.~\ref{cascades-barplots}. The actual values of multifractal measures corresponding to this figure are collected in Table. \ref{cascades-table}.
Note, that in the traditional measurement of the multifractal spread (see Fig.~\ref{artistic}), when $h_{obs}^{-}\geq h_b^{up}(-Q)$ and $h_{obs}^{+}\leq h_b^{down}(+Q)$ we have the following relations in terms of the edge $h^\pm$ values:
\begin{equation}
h^{-}_{obs}=h^{up}_{b}(-Q)+\Delta h^{-}
\end{equation}
\begin{equation}
h^{+}_{obs}=h^{down}_{b}(+Q)-\Delta h^{+}
\end{equation}
where $h^{\pm}_{obs}$ are the observed (biased) edge values of the multifractal profile and $\Delta h^{\mp}$ are increments above (below) the multifractal bias level.
The latter ones are assumed to be null if the edges $h^{\mp}_{obs}$ satisfy $h^{down}(\mp Q)<h^{\mp}_{obs}<h^{up}(\mp Q)$.

Therefore, one gets
\begin{equation}
\Delta h_{obs} = h^{-}_{obs} - h^{+}_{obs} = \Delta h_{b} + \Delta h_{unb}
\end{equation}
where $\Delta h_{b}=h^{up}_{b}(-Q)-h^{down}_{b}(+Q)$ is the bias level and $\Delta h_{unb}=\Delta h_{obs}-\Delta h_b$ is the unbiased spread of generalized Hurst exponents.

In case of synthetic multifractals, the new measure $\Delta_h$ gives results similar to unbiased multifractal spread $\Delta h_{unb}$ as seen from tabularized values in Table~\ref{cascades-table}. Both measures ($\Delta h_{unb}$ and $\Delta_h$) subtract the influence of finite size (FSE) and nonlinear transformation effects.

However, when a real complex system data are considered significant difference between new and standard measure can be observed.
We present an evolution of three stock exchange indices: S\&P500, Nikkei225 and NASDAQ in
Figs.~\ref{SP500},  \ref{Nikkei225} and \ref{NASDAQ} in two adjacent time windows.
Investigated examples are chosen in such a way, that each right hand side window contains an extreme event, that is: "Black Monday" of 19. October 1987 for S\&P500, fall crash in 2008 for Nikkei225 and the crash of 30. August 1995 for NASDAQ, while the left hand side window corresponds to proceeding "quiet" period on the stock market.
All data are investigated for primary price increments $\Delta x_i$ and their nonlinear transformations: 
absolute price increments $|\Delta x_i|$, 
quadratic increments $(\Delta x_i)^2$, 
absolute returns $|r_i|$ 
and moving volatilities 
$\mu_i = (1/s) \sum^i_{k=i-s+1} |r_i|$, 
$v_i = [(1/s) \sum^i_{k=i-s+1} (r_k-\langle r_k\rangle_s)^2]^{\frac{1}{2}}$, where $i=1,\ldots,1000$ and $s=21$ transaction days.
The properties of multifractal profiles $h(q)$ and multifractal bias coming from FSE and nonlinear transformation effects (red ribbon-like area) are presented in the first and last column in each figure.
The quantitative comparison of multifractal measures based on $\Delta h$ and $\Delta_h$ exponents calculated from these plots are collected in Tables 2--4 and plotted in Figs. \ref{SP500-barplots}--\ref{NASDAQ-barplots}.

Clear differences in multifractal features are visible between quiet and more dynamical periods (differences in shapes of $h(q)$ profiles), which makes them excellent candidates to confront the qualities of three measures $\Delta h_{obs}$, $\Delta h_{unb}$ and the proposed one $\Delta_h$.
No qualitative difference between both periods is visible for primary data of Nikkei225 index, when the standard measure $\Delta h_{unb}$ is applied.
S\&P500 and NASDAQ behave the other way round -- one observes richer multifractal structure for them in dynamical period.
For nonlinear transformations a strong non-monotonicity is mostly visible in dynamical period of all considered market indices leading in some cases ($|\Delta x_i|$ and $|r_i|$ for Nikkei225) to inversed behavior, where $h(-Q)<h(Q)$ (here $Q=15$).
Thus $\Delta h_{obs}<0$ and the traditional measure is not indicative in such cases.
The unbiased measure $\Delta h_{unb}$ for $|\Delta x_i|$ and $|r_i|$ of dynamical period for S\&P500, Nikkei225 and NASDAQ and $(\Delta x_i)^2$ just in case of Nikkei225 index causes also interpretation problems ($h_{obs}^-<h_b^{down}(-Q)$).
The above problematic situations are marked as cross-points in Figs. 8, 9, 10.
Unexpectedly, moving volatile transformations $\mu_i$ and $v_i$ for dynamical period of Nikkei225 index, show complete lack of unbiased multifractal effects contrary to other considered indices.

The new measure (see bottom panel in Figs. 8, 9, 10) maintains validity of conclusions stated about multifractal features found in traditional manner for primary series $\Delta x_i$.
But simultaneously, all prior problems with interpretation of multifractal properties of transformed data disappear. 
The interesting results are seen for two leading world indices S\&P500 and NASDAQ, particularly in dynamical zones.
The standard measure could not provide any quantitative result to be compared with the corresponding behavior in "quiet" zone (see Figs. \ref{SP500-barplots}--\ref{NASDAQ-barplots}, top panels).
The new measure makes it possible (see Figs. \ref{SP500-barplots}--\ref{NASDAQ-barplots}, bottom panels).
Moreover, one clearly notices that the index and its nonlinear transformations exhibit much richer multifractal content in a dynamical periods, then in the "quiet" zone.
Nikkei index is an exception, but also here a comparison between multifractal behaviors in two domains is possible only within the newly introduced measure (see Fig.~\ref{Nikkei225-barplots}).

\section{Conclusions}
This paper proposed the new method of measuring multifractal features in time series.
The method is based on global analysis of generalized Hurst exponents $h(q)$.
In the simplest case, when  multifractal bias is not considered, the new measure is defined as cumulated distance of the generalized Hurst exponents $h(q)$ from the main Hurst exponent $H$,
since $H$ is the only scaling exponent value one can refer to in monofractal case.
The generalisation to the case with multifractal bias from FSE and nonlinear transformations was also provided.

An example of synthetically generated multifractals with binomial cascades shows, that the new measure reproduces the standard spread of multifractal profile $\Delta h$, both before and after transformation.
However, when multifractal bias is taken into account, the new measure seems to be much more resistant to disturbances coming from different sources.

\begin{figure}[p]
\centering
\includegraphics[width=0.49\textwidth]{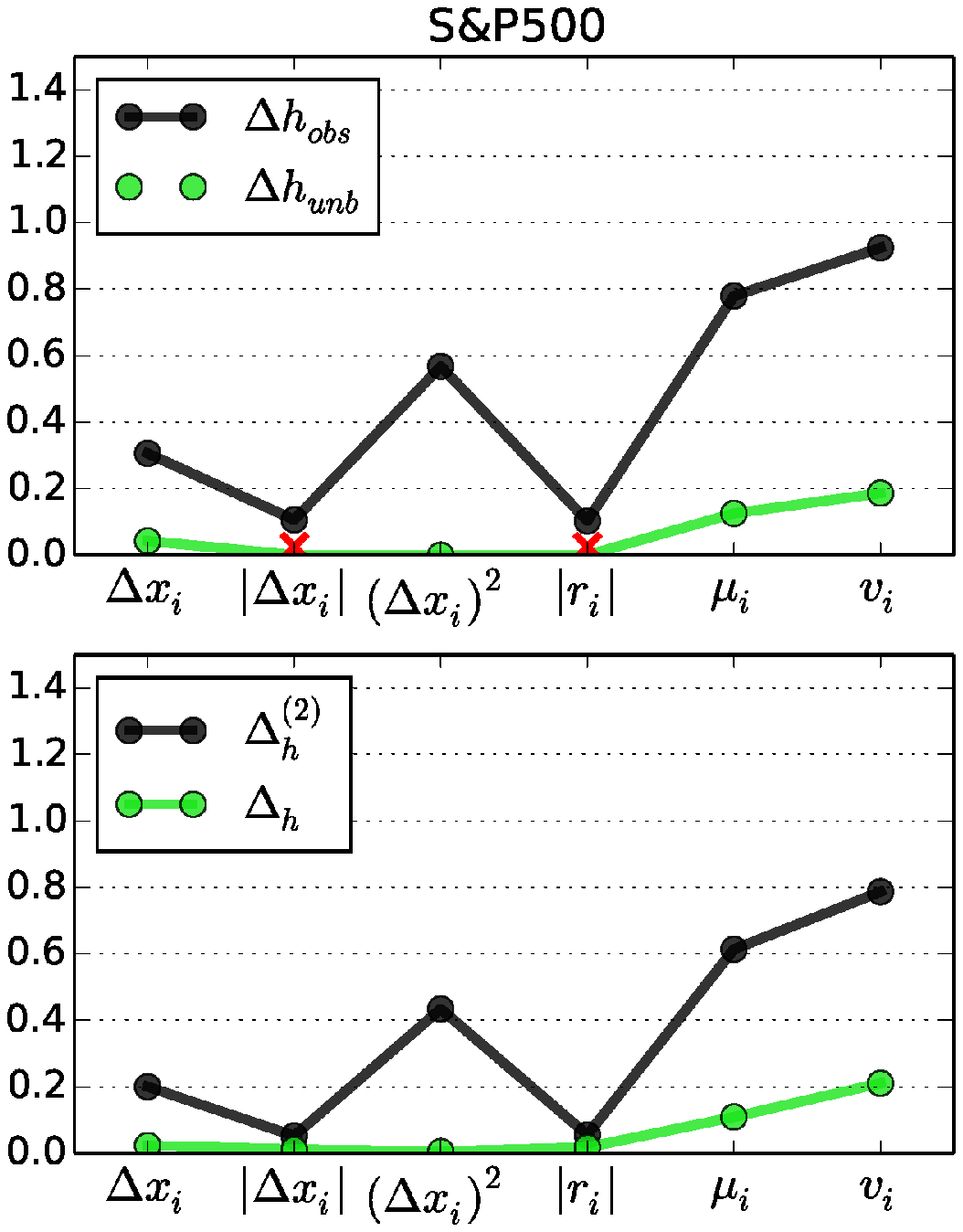}
\includegraphics[width=0.49\textwidth]{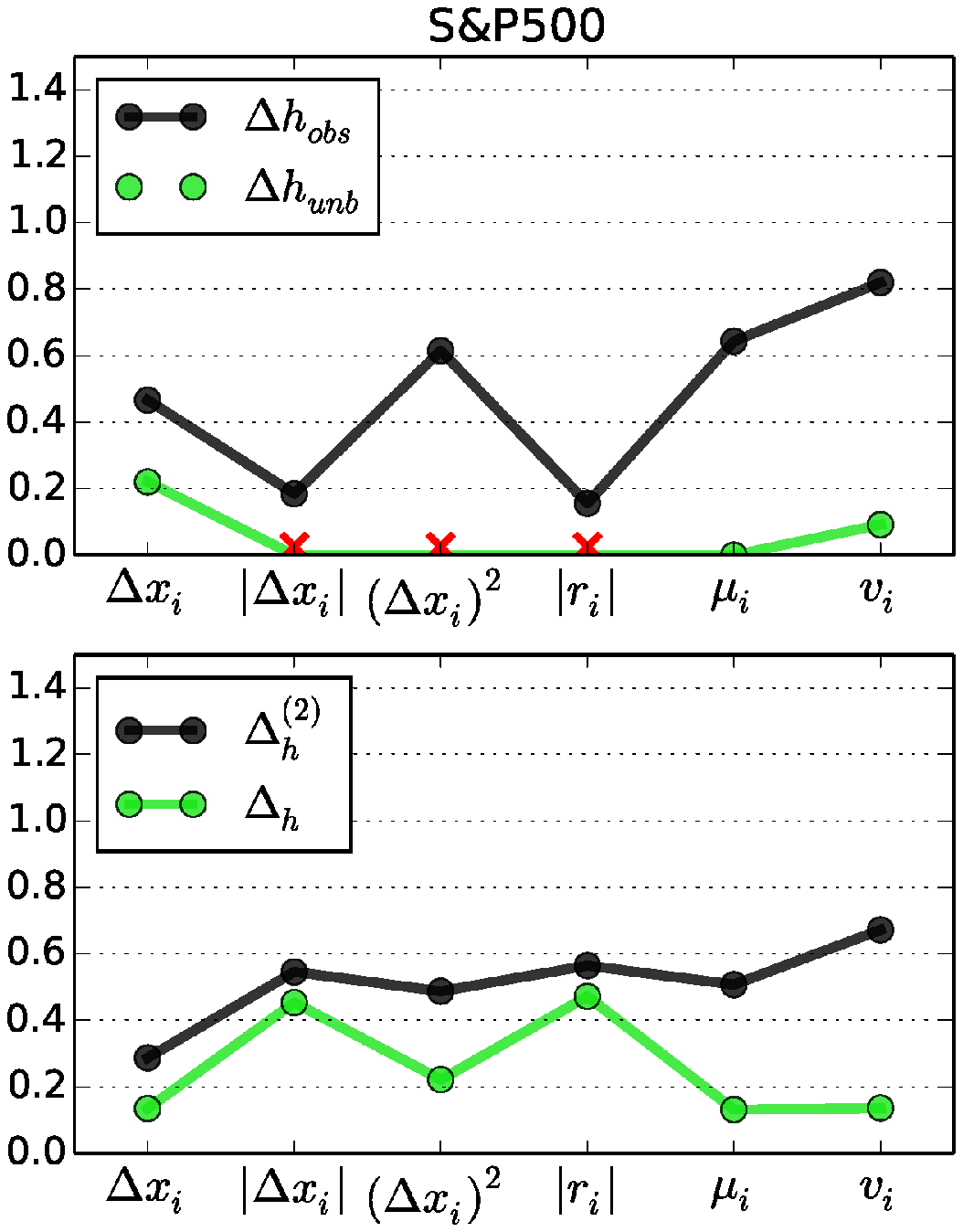}
\caption{Comparative study of multifractal measures for S\&P500 in two adjacent windows with "quiet" (left) and dynamical (right) behavior.
Results are presented in the same way as in Fig.~\ref{cascades-barplots}.
Cross-point in top plots correspond to cases, when traditional multifractal profile spreads $\Delta h_{obs}$ or $\Delta h_{unb}$ causes problems with interpretation ($h^-<h^+$ for $\Delta h_{obs}$ or $h^-<h_b^{down}$ or $h^+>h_b^{up}$ for $\Delta h_{unb}$). See also Fig.~\ref{artistic}.
}
\label{SP500-barplots}
\end{figure}

\begin{figure}[p]
\centering
\includegraphics[width=0.49\textwidth]{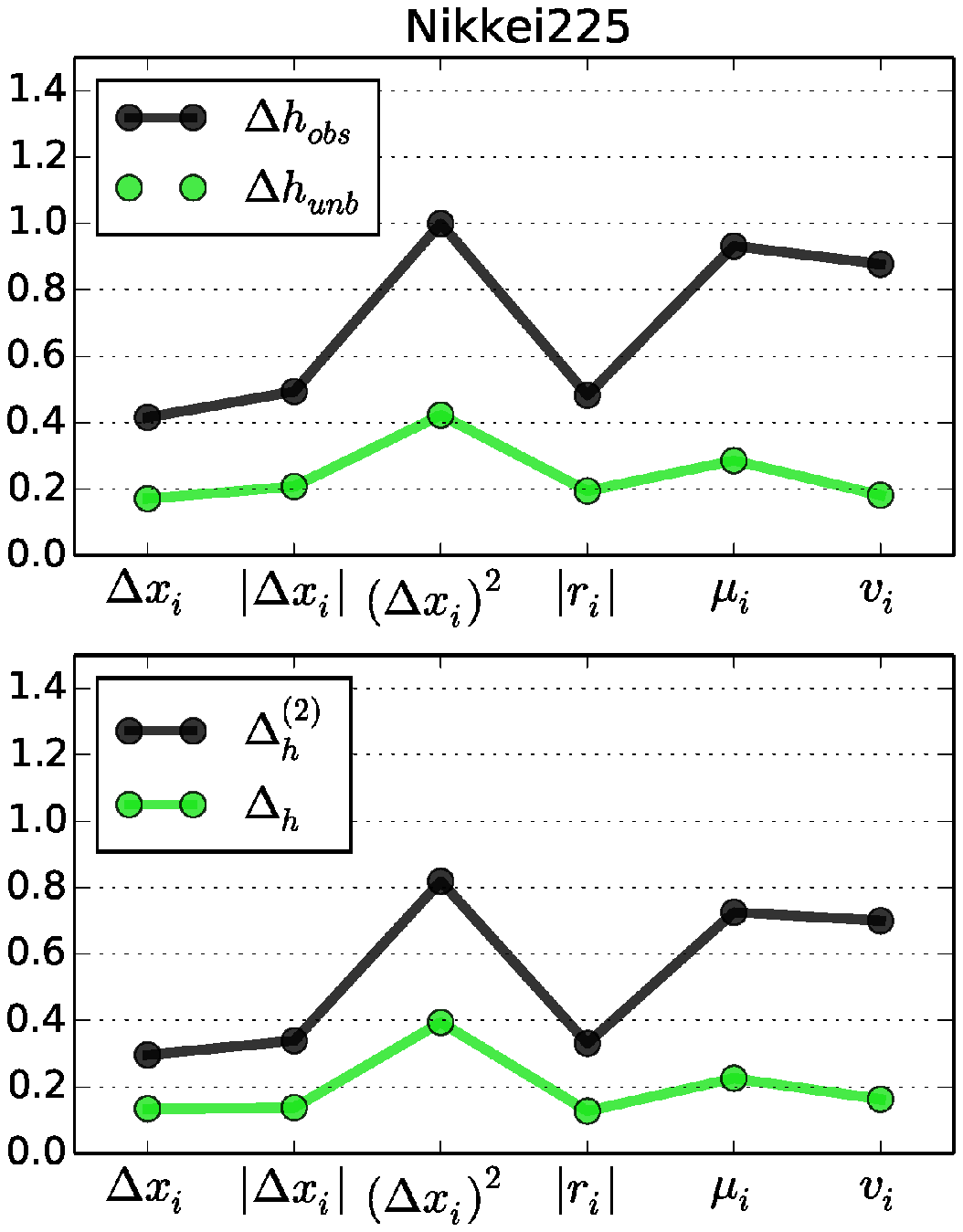}
\includegraphics[width=0.49\textwidth]{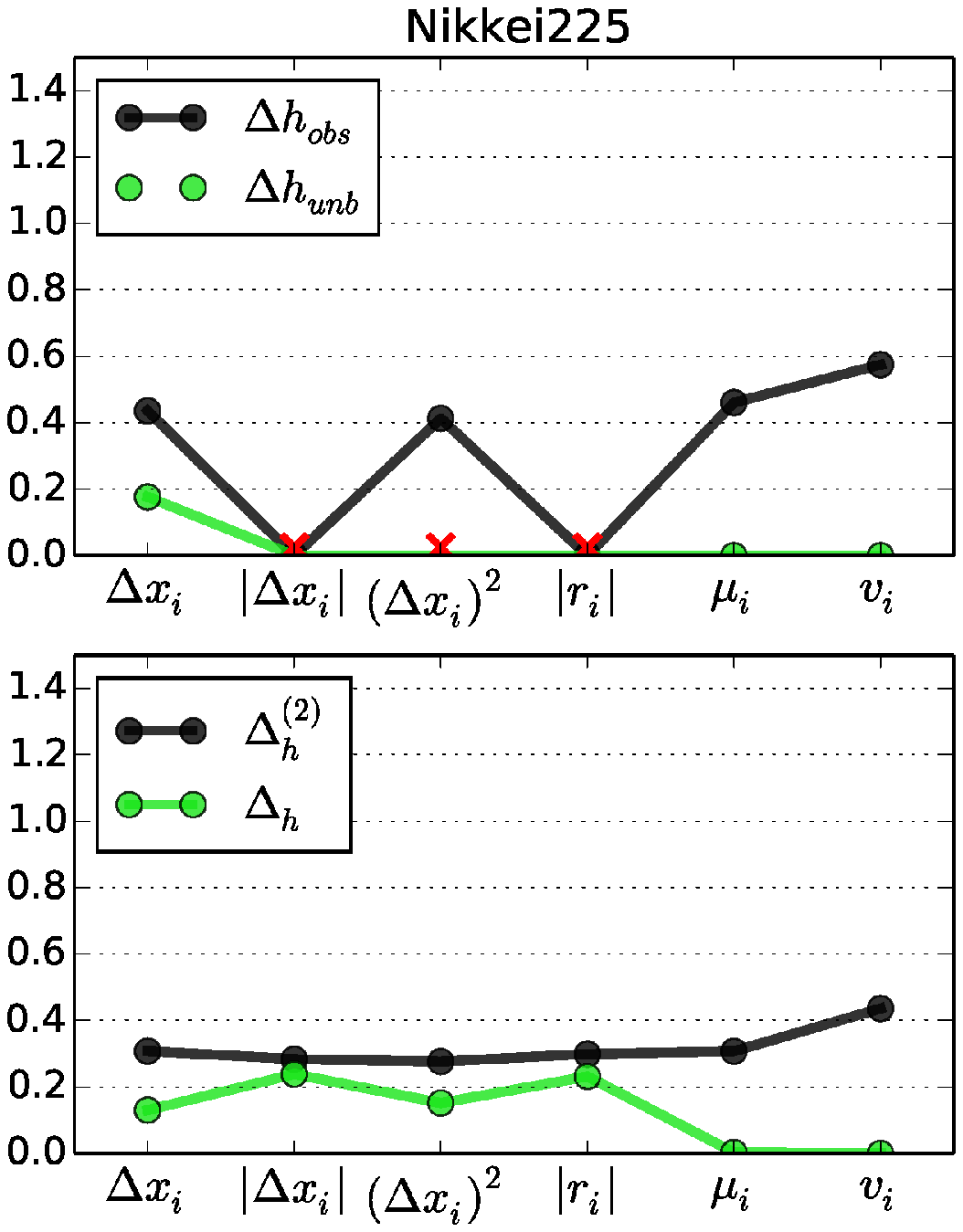}
\caption{As in Fig.~\ref{SP500-barplots}, but for Nikkei225 index.}
\label{Nikkei225-barplots}
\end{figure}

\begin{figure}[p]
\centering
\includegraphics[width=0.49\textwidth]{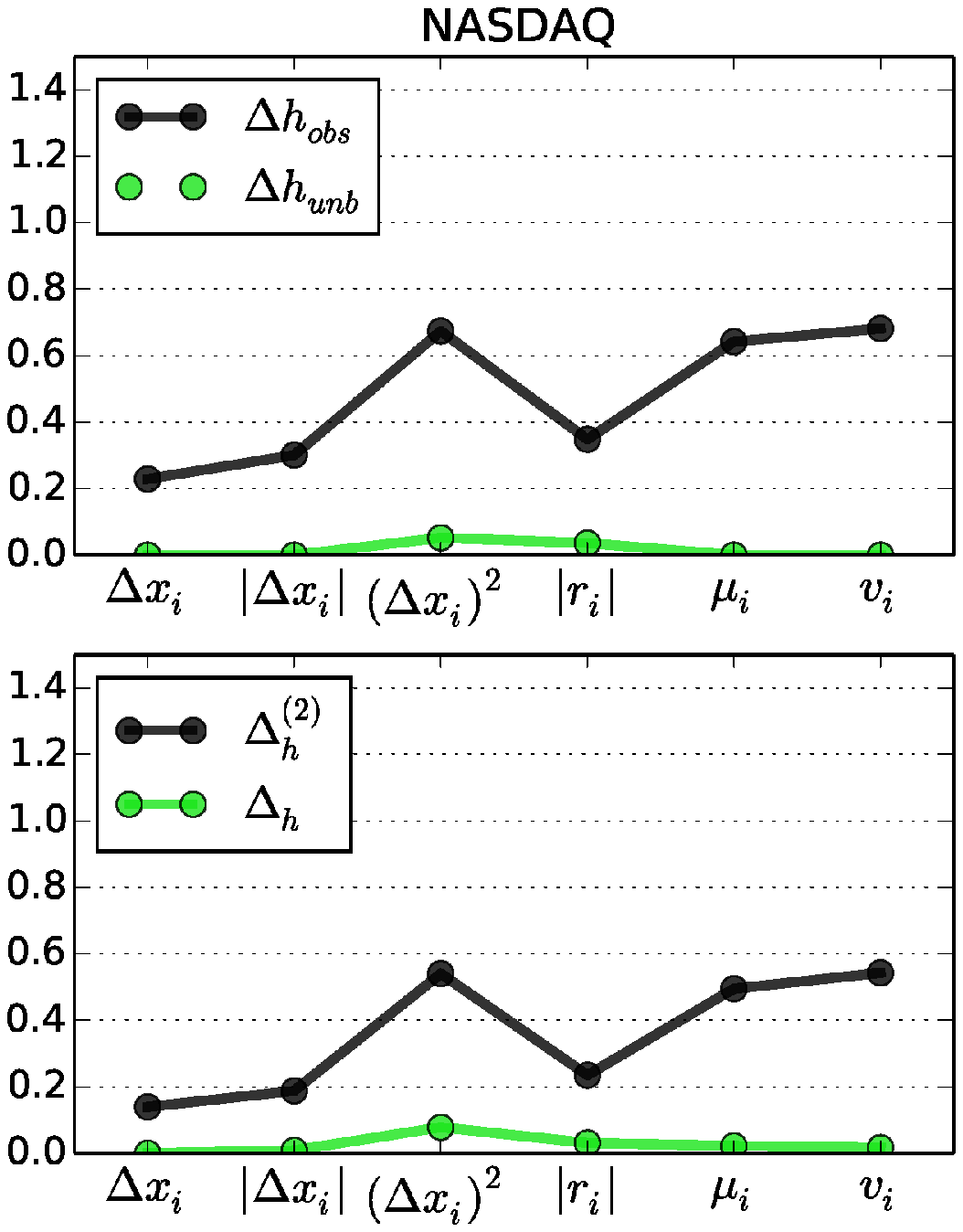}
\includegraphics[width=0.49\textwidth]{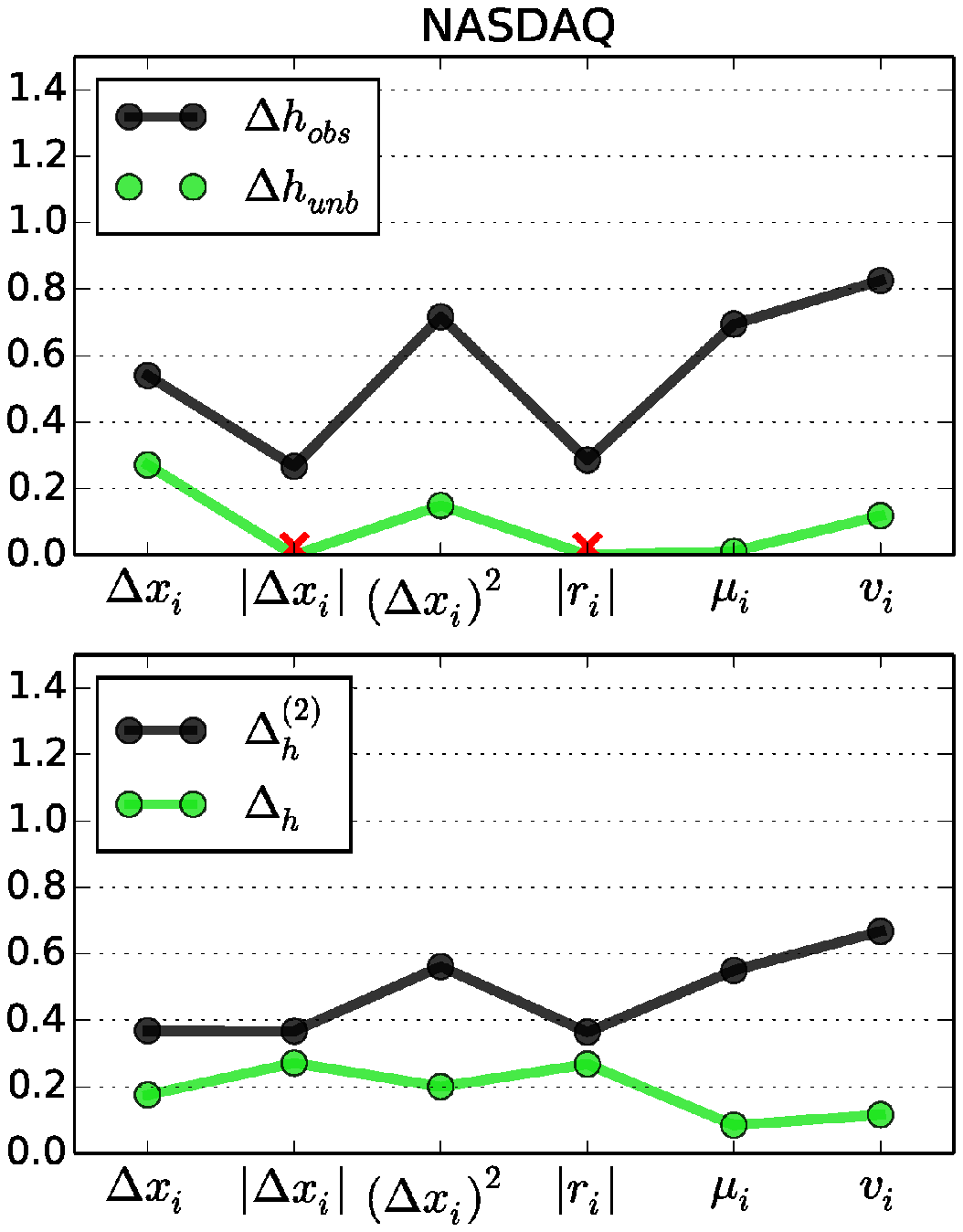}
\caption{As in Fig.~\ref{SP500-barplots}, but for NASDAQ index.}
\label{NASDAQ-barplots}
\end{figure}

We have presented the examples of real signals drawn from financial markets.
These examples prove, that the new technique of measuring multi-scaling features in time series is useful, especially in very dynamical periods of complex system evolution, when simpler predecessor ($\Delta h$) ceases to work properly (see transformations $|\Delta x_i|$, $(\Delta x_i)^2$ $|r_i|$, in right panels of Figs.~\ref{SP500-barplots} and~\ref{Nikkei225-barplots} and transformations $|\Delta x_i|$, $|r_i|$ in right panel of Fig.~\ref{NASDAQ-barplots}).
Obtained results seem to confirm, that whenever standard methods of detecting multifractality within MFDFA face problems, the newly proposed approach based on the scaling law from Eq. (12) detects it successfully. 

\clearpage

\begin{table}[h]
\begin{tabular}{||c||c|c|c|c|c|c||c|c|c|c|c|c||}
 & \multicolumn{6}{c||}{$a=0.65$} & \multicolumn{6}{c||}{$a=0.75$}\\
 & $\Delta x_i$ & $|\Delta x_i|$ & $(\Delta x_i)^2$ & $|r_i|$ & $\mu_i$ & $v_i$ & $\Delta x_i$ & $|\Delta x_i|$ & $(\Delta x_i)^2$ & $|r_i|$ & $\mu_i$ & $v_i$\\\hline
$\Delta h_{obs}$ & 0.82 & 0.82 & 1.72 & 0.82 & 0.98 & 0.91 & 1.49 & 1.47 & 3.08 & 1.43 & 1.49 & 1.45 \\
$\Delta h_{unb}$ & 0.73 & 0.78 & 1.36 & 0.64 & 0.17 & 0.40 & 1.41 & 1.38 & 2.69 & 1.29 & 0.68 & 0.91 \\
$\Delta_h^{(2)}$ & 0.61 & 0.68 & 1.49 & 0.66 & 0.76 & 0.69 & 1.27 & 1.31 & 2.86 & 1.26 & 1.26 & 1.26 \\
$\Delta_h$       & 0.60 & 0.68 & 1.28 & 0.60 & 0.16 & 0.35 & 1.26 & 1.30 & 2.66 & 1.24 & 0.41 & 0.56 \\
\end{tabular}
\caption{
Results of multifractal measures for synthetic data series generated from random binomial cascade algorithm
 corresponding to plots in  Fig.~\ref{cascades-barplots}.
}
\label{cascades-table}
\end{table}

\begin{table}[h]
\begin{tabular}{||c||c|c|c|c|c|c||c|c|c|c|c|c||}
 & \multicolumn{6}{c||}{"quiet" period} & \multicolumn{6}{c||}{dynamical period}\\
 & $\Delta x_i$ & $|\Delta x_i|$ & $(\Delta x_i)^2$ & $|r_i|$ & $\mu_i$ & $v_i$ & $\Delta x_i$ & $|\Delta x_i|$ & $(\Delta x_i)^2$ & $|r_i|$ & $\mu_i$ & $v_i$\\\hline
$\Delta h_{obs}$ & 0.31 & 0.11     & 0.57 & 0.10     & 0.78 & 0.92 & 0.47 & 0.18     & 0.62     & 0.16     & 0.64 & 0.82\\
$\Delta h_{unb}$ & 0.04 & $\times$ & 0.00 & $\times$ & 0.13 & 0.19 & 0.22 & $\times$ & $\times$ & $\times$ & 0.00 & 0.09\\
$\Delta_h^{(2)}$ & 0.20 & 0.05     & 0.43 & 0.06     & 0.61 & 0.79 & 0.29 & 0.55     & 0.49     & 0.56     & 0.51 & 0.67\\
$\Delta_h$       & 0.02 & 0.01     & 0.01 & 0.02     & 0.11 & 0.21 & 0.13 & 0.45     & 0.22     & 0.47     & 0.13 & 0.14\\
\end{tabular}
\caption{
Results of multifractal measures for S\&P500 index
 corresponding to plots in  Fig.~\ref{SP500-barplots}.
Crosses describe cases, when traditional multifractal profile spread $\Delta h_{unb}$ causes problems with interpretation ($h^-<h_b^{down}$ or $h^+>h_b^{up}$).
}
\end{table}

\begin{table}[h]
\begin{tabular}{||c||c|c|c|c|c|c||c|c|c|c|c|c||}
 & \multicolumn{6}{c||}{"quiet" period} & \multicolumn{6}{c||}{dynamical period}\\
 & $\Delta x_i$ & $|\Delta x_i|$ & $(\Delta x_i)^2$ & $|r_i|$ & $\mu_i$ & $v_i$ & $\Delta x_i$ & $|\Delta x_i|$ & $(\Delta x_i)^2$ & $|r_i|$ & $\mu_i$ & $v_i$\\\hline
$\Delta h_{obs}$ & 0.42 & 0.49 & 1.00 & 0.48 & 0.93 & 0.88 & 0.44 & $\times$ & 0.41     & $\times$ & 0.46 & 0.58\\
$\Delta h_{unb}$ & 0.17 & 0.21 & 0.42 & 0.19 & 0.29 & 0.18 & 0.18 & $\times$ & $\times$ & $\times$ & 0.00 & 0.00\\
$\Delta_h^{(2)}$ & 0.30 & 0.34 & 0.82 & 0.33 & 0.73 & 0.70 & 0.31 & 0.28     & 0.28     & 0.30     & 0.31 & 0.44\\
$\Delta_h$       & 0.13 & 0.14 & 0.39 & 0.13 & 0.23 & 0.16 & 0.13 & 0.24     & 0.15     & 0.23     & 0.00 & 0.00\\
\end{tabular}
\caption{
Results of multifractal measures for Nikkei225 index
 corresponding to plots in  Fig.~\ref{Nikkei225-barplots}.
Crosses describe cases, when traditional multifractal profile spreads $\Delta h_{obs}$ or $\Delta h_{unb}$ causes problems with interpretation ($h^-<h^+$ for $\Delta h_{obs}$ or $h^-<h_b^{down}$ or $h^+>h_b^{up}$ for $\Delta h_{unb}$).
}
\end{table}

\begin{table}[!h]
\begin{tabular}{||c||c|c|c|c|c|c||c|c|c|c|c|c||}
 & \multicolumn{6}{c||}{"quiet" period} & \multicolumn{6}{c||}{dynamical period}\\
 & $\Delta x_i$ & $|\Delta x_i|$ & $(\Delta x_i)^2$ & $|r_i|$ & $\mu_i$ & $v_i$ & $\Delta x_i$ & $|\Delta x_i|$ & $(\Delta x_i)^2$ & $|r_i|$ & $\mu_i$ & $v_i$\\\hline
$\Delta h_{obs}$ & 0.23 & 0.30 & 0.67 & 0.35 & 0.64 & 0.68 & 0.54 & 0.27     & 0.72 & 0.29     & 0.69 & 0.83\\
$\Delta h_{unb}$ & 0.00 & 0.00 & 0.05 & 0.04 & 0.00 & 0.00 & 0.27 & $\times$ & 0.15 & $\times$ & 0.01 & 0.12\\
$\Delta_h^{(2)}$ & 0.14 & 0.19 & 0.54 & 0.24 & 0.50 & 0.54 & 0.37 & 0.37     & 0.56 & 0.36     & 0.55 & 0.67\\
$\Delta_h$       & 0.00 & 0.01 & 0.08 & 0.03 & 0.02 & 0.02 & 0.18 & 0.27     & 0.20 & 0.27     & 0.08 & 0.12\\
\end{tabular}
\caption{
Results of multifractal measures for NASDAQ index
 corresponding to plots in  Fig.~\ref{NASDAQ-barplots}.
Crosses describe cases, when traditional multifractal profile spread $\Delta h_{unb}$ causes problems with interpretation ($h^-<h_b^{down}$ or $h^+>h_b^{up}$).
}
\end{table}
\clearpage

\end{document}